\documentclass{webofc}
\usepackage[varg]{txfonts}   % Web of Conferences font
\usepackage{hyperref}
\usepackage{upgreek}
\usepackage{url}
\usepackage{lineno}
\usepackage{orcidlink}
\hypersetup{hidelinks}

\hypersetup{colorlinks=true,citecolor=blue,urlcolor=blue,linkcolor=blue}
\begin{document}
\title{The ITS3 detector and physics reach of the LS3 ALICE Upgrade}

\author{\firstname{Chun-Zheng}
\lastname{Wang\orcidlink{0000-0001-5383-0970} (for the ALICE Collaboration)}\inst{1,2}\fnsep\thanks{\email{chunzheng.wang@cern.ch}}
}

\institute{Key Laboratory of Nuclear Physics and Ion-beam Application (MOE), Institute of Modern Physics, Fudan University, 200433, Shanghai, China
\and
        Key Laboratory of Quark \& Lepton Physics (MOE) and Institute of Particle Physics, Central China Normal University, 430079, Wuhan, China
}

\abstract{During Large Hadron Collider (LHC) Long Shutdown 3 (LS3) (2026-28), the ALICE experiment is replacing its inner-most three tracking layers by a new detector, Inner Tracking System 3. It will be based on newly developed wafer-scale monolithic active pixel sensors, which are bent into truly cylindrical layers and held in place by light mechanics made from carbon foam. Unprecedented low values of material budget (per layer) and closeness to interaction point (19 mm) lead to a factor two improvement in pointing resolutions from very low $p_\text{T}$ (O(100MeV/$c$)), achieving, for example, 20 ${\upmu}$m and 15 ${\upmu}$m in the transversal and longitudinal directions, respectively, for 1 GeV/c primary charged pions.
After a successful R\&D phase 2019-2023, which demonstrated the feasibility of this innovational detector, the final sensor and mechanics are being developed right now. This contribution will briefly review the conceptual design and the main R\&D achievements, as well as the current activities and road to completion and installation. It concludes with a projection of the improved physics performance, in particular for heavy-flavour hadrons, as well as for thermal dielectrons, that will come into reach with this new detector installed.
}
\maketitle
\vspace{-0.4cm}
\section{Introduction}
\label{intro}
ALICE (A Large Ion Collider Experiment) is a detector at LHC specifically designed to investigate high-energy heavy-ion collisions, which could lead to the formation of the quark--gluon plasma, a colour-deconfined state of the matter. To achieve this, ALICE is equipped with multiple detectors, including the ITS, which plays a crucial role in determining the interaction and decay vertices and precisely tracking particles with low momentum and angular resolution in conjunction with the TPC and other detectors. The ITS was upgraded during LHC LS2 to the ITS2, which is constructed using ALPIDE chips based on 180 nm CMOS technology \cite{ALPIDE}.
% , fabricated with a 180 nm process, and provides a substantial 10 m² active silicon area containing approximately 12.5 billion pixels. 

% In the ITS2’s inner barrel, each layer is composed of multiple staves, with each stave consisting of a carbon support, a cooling plate, ALPIDE chips, and a flexible printed circuit (FPC). The staves overlap to ensure complete coverage; however, these overlaps introduce non-sensitive materials, such as water and Kapton, leading to uneven material distribution in the azimuthal angle.

\begin{figure*}
\centering
\includegraphics[width=10cm,clip]{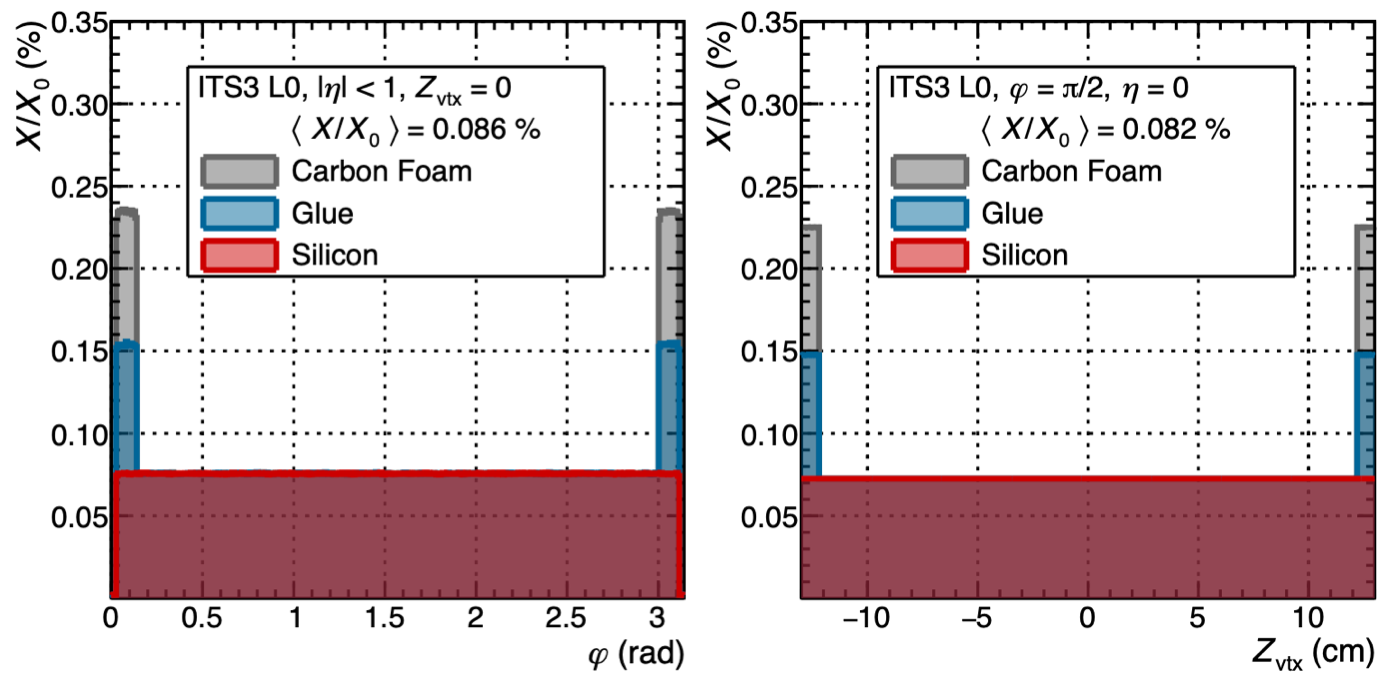}
\vspace{-0.4cm}
\caption{Material budget for tracks of particles originating from the interaction point as a function of as a function of the azimuthal angle $\phi$ (left panel) and the coordinate of the interaction vertex along the beam axis $Z_{\mathrm{vtx}}$ (right panel)}
\vspace{-0.4cm}
\label{fig-MBD}       % Give a unique label
\end{figure*}

With advances in chip manufacturing technology, wafer-scale chips can now be produced using stitching techniques. This development has opened new avenues for the next upgrade of the ITS. After a successful R\&D phase 2019-23, a next-generation ITS3 design has been proposed by ALICE \cite{TDR}, aimed at replacing the innermost three tracking layers of ITS2 during LHC LS3 (2026-28). ITS3 features three layers of wafer-scale chips manufactured using stitching technology, and these chips will be bent into a half-cylinder shape. The chips, based on 65 nm CMOS MAPS technology, with a power density below 40 mW/cm$^2$, allow for air cooling to replace the water cooling system used in ITS2. These improvements will significantly reduce the material budget from 0.35\% $X_{0}$ to 0.07\% $X_{0}$ per layer from ITS2 to ITS3. Additionally, the material distribution will become more uniform as Fig. \ref{fig-MBD} illustrates. Furthermore, the beam pipe’s radius will be reduced from 18.2 mm to 16.0 mm, leading to a decrease in the radius of layer 0 of ITS3 from approximately 24.0 mm to 19.0 mm.

\section{Chip design and characterization}
\label{sec-1}
Supporting this ambitious design, the chip development for ITS3 involves multiple phases. In the first phase, the MLR1 phase, three types of chips—Analogue Pixel Test Structure (APTS), Digital Pixel Test Structure (DPTS), and Circuit Exploratoire 65 nm (CE65). Through this phase, the 65 nm process chips were thoroughly studied, including the confirmation of radiation hardness. Even when subjected to 100 times the radiation dose required for ITS3 (10 kGy and $10^{13}$ 1MeV $\text{n}_{\text{eq}}$ cm$^2$), the detection efficiency of the chips remained sufficiently high, and the spatial resolution was almost unaffected. For more detailed results, please refer to Ref. \cite{MLR1}.

\begin{figure*}
\centering
\includegraphics[width=11cm,clip]{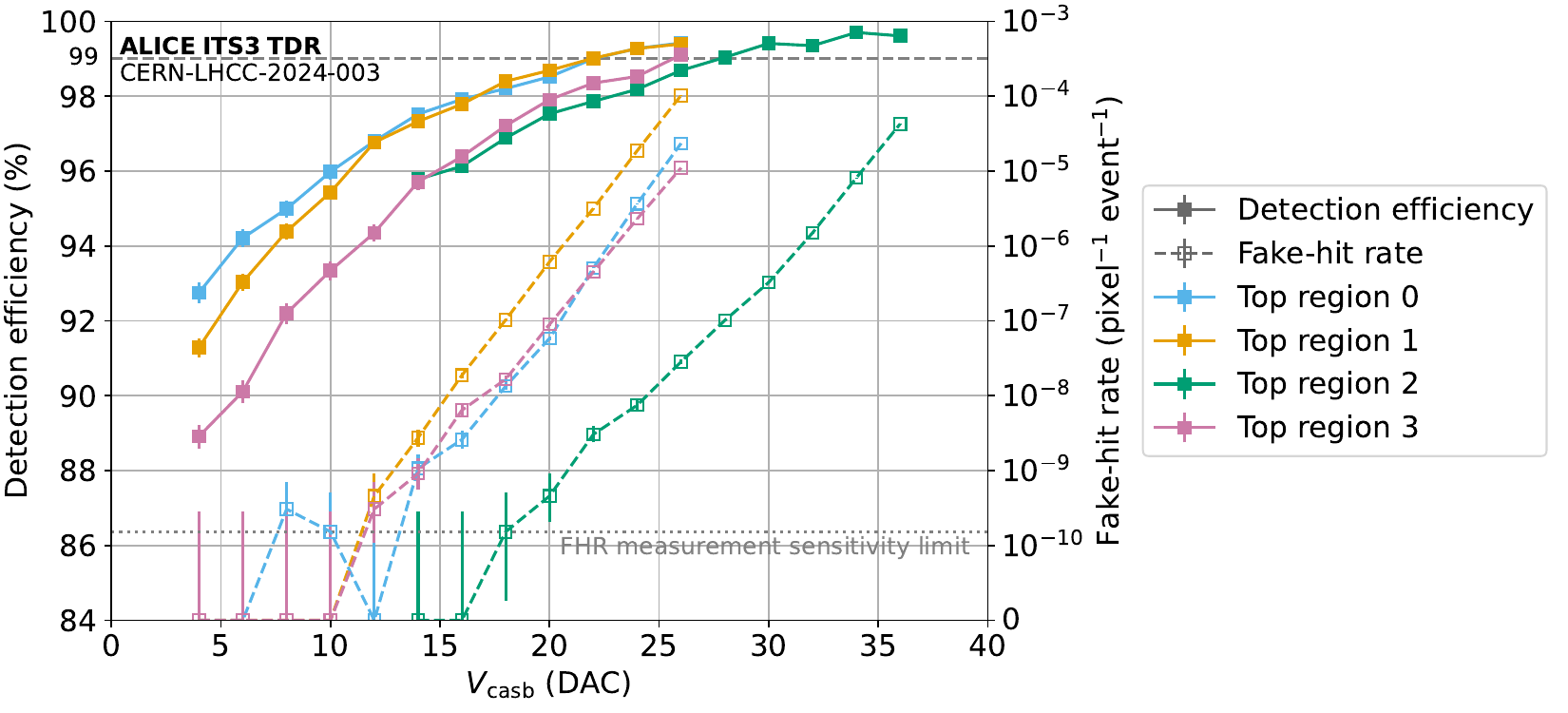}
\vspace{-0.4cm}
\caption{Detection efficiency and fake-hit rate as a function of $V_\text{casb}$ (inversely proportional to the threshold in number of electrons) measured for MOSS at the CERN PS.}
\vspace{-0.4cm}
\label{fig-MOSSEff}       % Give a unique label
\end{figure*}

In the ER1 phase, ALICE designed two types of wafer-scale chips: MOnolithic Stitched sensor (MOSS) and MOnolithic Stitched sensor with Timing (MOST). The MOSS chip is the first stitched chip used for high-energy physics, with a length of 259 mm and a width of 14 mm. The MOSS chip is segmented into a left end cap (LEC), a right cap (REC), and 10 repeated sensor units (RSU), with each RSU containing a top and a bottom half-unit. The top half-unit is divided into four pixel regions, where each pixel has a width of 22.5${\upmu}$m, and there are 256 $\times$ 256 pixels per region, with slight design variations among the regions. The bottom half-unit is also divided into four regions, each with a pixel width of 18${\upmu}$m and 320 $\times$ 320 pixels. Fig. \ref{fig-MOSSEff} shows the beam test results for the MOSS chip at CERN PS, which indicate that the MOSS chip achieves a detection efficiency of over 99\% with a very low fake-hit rate, below $10^{-6}$ pixel$^{-1}$ event$^{-1}$ for region 2 when operated at an optimal working point. More detailed description can be found in the ITS3 TDR\cite{TDR}.

The successful design of the MOSS chip has confirmed the effectiveness of using wafer-scale chips for particle detection, and provided the guidance for the design and fabrication of ITS3 chips in the subsequent ER2 phase.

\section{Physics performance and physics reach of ITS3}

Using the $\mathrm{O}^2$ framework \cite{O2} developed by ALICE for the reconstruction, calibration, and simulation of the ALICE experiment for LHC Run 3 and 4, the impact parameter resolution of ITS3 has been estimated using two simulation methods: full simulation and fast analytic tool (FAT), both yielding consistent results. Compared to the ITS2, the ITS3 is expected to achieve a twofold improvement in impact parameter resolution \cite{SPP}. This will greatly benefit numerous analyses, such as studies of the participation of heavy-flavour particles in the collective motions of the system created in heavy-ion collisions

\begin{figure*}
\centering
\includegraphics[width=11cm,clip]{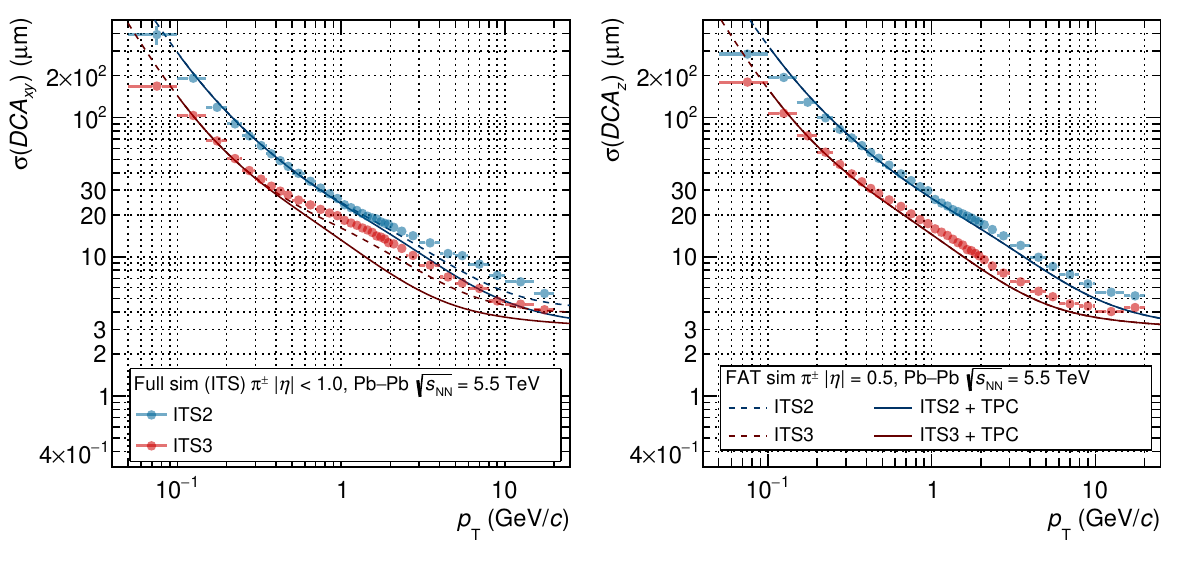}
\vspace{-0.4cm}
\caption{Impact parameter resolution in the $r\phi$ (left panel) and longitudinal (right panel) direction for primary charged pions with $|\eta| < 1$ as a function of the track $p_{\text{T}}$ for ITS2 and ITS3 detectors from the O$^2$ simulation}
\vspace{-0.4cm}
\label{fig-sig}       % Give a unique label
\end{figure*}

\begin{figure*}
\centering
\includegraphics[width=11.5cm,clip]{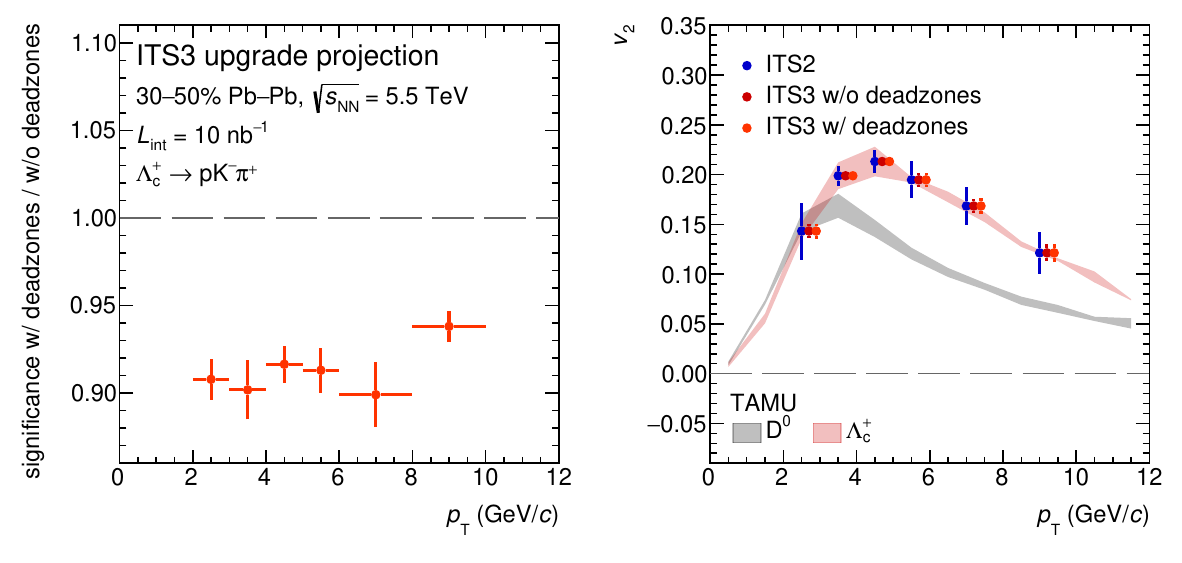}
\vspace{-0.4cm}
\caption{Left: Ratio between the expected significance with or without deadzones. Right panel: Comparison of the expected performance for the measurement of the $\Lambda_\text{c}$ elliptic flow with the ITS2 and the ITS3 with and without deadzones.}
\vspace{-0.4cm}
\label{fig-Lc}       % Give a unique label
\end{figure*}

The production of heavy flavor particles in the QGP mainly arises from two mechanisms: fragmentation, where the transition from heavy quark to hadron occurs via the emission of a parton shower, resulting in a hadron that carries a fraction of the original parton momentum, and coalescence, where partons close in phase space can recombine. These two mechanisms contribute differently to the elliptic flow coefficient \cite{FLOW} of heavy-flavor particles. The performance of the ITS3 in reconstructing heavy-flavor particles is expected to be significantly better compared to the ITS2. As shown in the left panel of Fig. \ref{fig-Lc}, although the dead zones have a minor impact, the reconstruction remains extremely precise, leading to accurate flow calculations. The right panel of Fig. \ref{fig-Lc} further demonstrates, according to the TAMU model \cite{TAMU}, that in the $p_\text{T}$ > 4 GeV/c region, there will be a significant difference in the flow of $\Lambda_\text{c}^{+}$ and $\text{D}^{0}$ particles. With ITS3, the statistical uncertainty in the $v_2$ measurement for $\Lambda_\text{c}^{+}$ can be reduced by up to a factor of 4. This study utilizing ITS3 will further constrain the modeling of charm diffusion and hadronization in the QGP. More details on the physics performance can be found in Ref. \cite{SPP}.

% Figure  shows the signal-to-background ratio, significance of the reconstructed $\Lambda_\text{c}$ particles using ITS2 and ITS3, and the difference in significance when considering the dead zone of the ITS3 chip. The signal-to-background ratio of ITS3 is expected to improve by a factor of 10, and the significance by a factor of 4. The impact of the dead zone is minimal compared to the overall improvements.

% \begin{figure*}
% \centering
% \includegraphics[width=11.5cm,clip]{sig.png}
% \caption{Performance of $\Lambda_\text{c} \rightarrow pK^{-}\pi^{+}$ reconstruction in central Pb–Pb collisions at $\sqrt{s_{\text{NN}}} = 5.5
% $ TeV: statistical significance (left) and S/B ratio (right) as a function of $p_{T}$.}
% \label{fig-sig}       % Give a unique label
% \end{figure*}

%
% BibTeX or Biber users please use (the style is already called in the class, ensure that the "woc.bst" style is in your local directory)
% \bibliography{your_bib_file} % Replace "your_bib_file" with the actual name of your .bib file

\section{Summary}

The ITS3, a bent wafer-scale monolithic pixel detector, is set to replace the ITS2 inner barrel during LHC LS3. Characterization results from each phase of chip design and production have confirmed that the bent 65nm process wafer-scale chips fully meet the requirements of ITS3, including radiation hardness, spatial resolution, and detection efficiency. The ITS3 project remains on track for installation as planned. This upgrade, offering a twofold improvement in spatial resolution compared to ITS2, will significantly enhance various analyses, such as heavy-flavor collectivity and  thermal dielectrons measurement.

%\section*{ACKNOWLEDGMENTS}
The author is supported in part by the National Key Research and Development Program of China (Nos. 2022YFA1602103, 2018YFE0104600), the National Natural Science Foundation of China (No. 12322508) and the Science and Technology Commission of Shanghai Municipality (No. 23590780100).

\end{document}